\documentclass[aps,prd,twocolumn,eqsecnum,amssymb,amsmath,showpacs,a4paper,superscriptaddress,nofootinbib,longbibliography]{revtex4-2}
\usepackage{graphicx}
\usepackage{amsfonts}
\usepackage{amsmath}
\usepackage{hyperref}
\usepackage{adjustbox}
\usepackage{float} 
\usepackage[normalem]{ulem}
\usepackage[nameinlink]{cleveref}
\usepackage[table,xcdraw]{xcolor}
\Crefname{figure}{Fig.}{Figs.}

\begin{document}

\title{Superradiant scattering by rotating black-bounce black holes}

\author{Pedro Henrique Croti Siqueira}
 \email{pedro.croti@ufabc.edu.br}
 \affiliation{Centro de Matem\'atica, Computa\c c\~ao e Cogni\c c\~ao, Universidade Federal do ABC (UFABC), 09210-170 Santo Andr\'e, S\~ao Paulo, Brazil}
 \affiliation{Centro de Ci\^encias Naturais e Humanas, Universidade Federal do ABC (UFABC), 09210-170 Santo Andr\'e, S\~ao Paulo, Brazil}
\author{Maur\'icio Richartz}
 \email{mauricio.richartz@ufabc.edu.br}
\affiliation{Centro de Matem\'atica, Computa\c c\~ao e Cogni\c c\~ao, Universidade Federal do ABC (UFABC), 09210-170 Santo Andr\'e, S\~ao Paulo, Brazil}

\begin{abstract}
\noindent

We investigate superradiant scattering off a rotating regular black hole described by a black-bounce metric which generalizes the Kerr spacetime of mass $M$ and specific angular momentum $a$ through a regularization parameter $p$ and two deformation exponents $(k,n)$.
Focusing on massless $(\ell,m)=(1,1)$ scalar modes, we explore the parameter space and compute amplification factors by numerically integrating the separated radial Klein–Gordon equation. We track the peak amplification and the corresponding frequency across the $(a/M,p/M)$ parameter space for several combinations of $k$ and $n$. We find that increasing $n$ systematically enhances superradiance, whereas increasing $k$ tends to suppress it. In particular, certain configurations can lead to amplification factors that significantly exceed the maximum amplification found for standard Kerr black holes.
\end{abstract}

\maketitle

\section{Introduction}

The detection of gravitational waves by observatories such as LIGO, Virgo, and KAGRA has opened new perspectives for testing the strong field regime of gravity~\cite{Dreyer:2003bv,Berti:2005ys,Berti:2007zu,Berti:2016lat,Brito:2018rfr,Ota:2019bzl}. In this context, the so-called Kerr hypothesis, which posits that astrophysical black holes are accurately described by the Kerr metric, has been extensively scrutinized~\cite{Bambi:2011mj,Herdeiro:2022yle}. Although current observations are largely consistent with this paradigm, uncertainties in the inferred masses and spins of compact objects leave room for deviations from the Kerr geometry and for alternative theories of gravity~\cite{KZ;2016,Santos:2024pfa,LIGOScientific:2025rid,Siegel:2025xgb,Zhao:2025mwq,Shaikh:2021yux}, thus motivating the investigation of more general, possibly regular, black hole solutions~\cite{Carballo-Rubio:2025fnc}.

Regular black holes possess an event horizon while remaining free from curvature singularities~\cite{Carballo-Rubio:2019nel,Carballo-Rubio:2019fnb}. The existence of regular black holes inevitably requires a violation of the strong energy condition in general relativity~\cite{Lobo:2020ffi,Zaslavskii:2010qz}. The first example of such an object was introduced by Bardeen~\cite{bardeen1968proceedings} and, since then, a variety of regular geometries have been idealized within classical or semiclassical frameworks~\cite{Hayward2006,Roman1983,Dymnikova1992,Fan:2016hvf,Ghosh:2014pba,Ansoldi:2008jw,Bronnikov:2005gm,Bronnikov:2006fu,Uchikata:2012zs,Balart:2014jia,Balart:2014cga,Culetu:2015cna}. Among the most recent proposals is the black-bounce spacetime introduced by Simpson and Visser~\cite{Simpson_2019,Simpson:2019cer,Lobo:2020ffi}, which describes either a regular black hole or a traversable wormhole depending on the values of the parameters that define the metric. In recent years, numerous studies have explored the structure of black-bounce spacetimes and their properties~\cite{Rodrigues:2023vtm,Canate:2022gpy,Rodrigues:2022rfj,Huang:2019arj,Bronnikov:2023aya,Lobo:2020ffi,Rodrigues:2025plw,Alencar:2024yvh,Pereira:2025xnw,Pereira:2024rtv,Pereira:2024gsl,Pereira:2023lck,Lima:2023jtl,Santos:2025xbk,Moreira:2025lip,Crispim:2025cql,LimaJunior:2025uyj,Bragado:2025jrg,Duran-Cabaces:2025sly,Pereira:2025fvg,Calza:2024xdh,Vagnozzi:2022moj,Calza:2025yfm,Barrientos:2022avi,Calza:2022ioe}.

Although the original Simpson-Visser spacetime is static, subsequent rotating generalizations have also attracted significant attention~\cite{Mazza:2021rgq,Xu:2021lff,Yang:2022xxh, Franzin:2022iai,Viththani:2024map,Khoo:2025qjc,Saha:2025hqm,Carneiro:2024ysn,Pedrotti:2024znu,Yang:2023agi,Franzin:2022wai,Kamenshchik:2023woo}. These geometries extend the classical Kerr solution by incorporating a regular core, while retaining key features such as an ergoregion. In particular, rotating regular black holes can support wave scattering phenomena similar to those found in the Kerr spacetime.
Among these phenomena, superradiance stands out as a particularly remarkable effect in which low-frequency waves interacting with the rotating geometry can be amplified~\cite{1971JETPL..14..180Z,1972JETP...35.1085Z,Brito:2015oca}. Superradiant amplification typically occurs when the wave frequency $\omega$ and the azimuthal number $m$ satisfy
\begin{equation}
0 < \omega < m \Omega_H,
\end{equation}
where $\Omega_H$ is the angular velocity of the event horizon. The process extracts rotational energy from the black hole and may lead to instabilities when the field is massive~\cite{Dolan2007,Dolan:2012yt,beyer2,Furuhashi:2004jk}.

In the present work, we explore the occurrence of superradiance in the rotating black-bounce geometry introduced in Ref.~\cite{Xu:2021lff}, which generalizes the black-bounce geometries of Refs.~\cite{Simpson_2019,Mazza:2021rgq}. By analyzing the scattering of massless scalar fields, we investigate how the deformation parameters of the metric affect the efficiency of superradiant amplification, specially in the large spin regime. In particular, our goal is to determine the peak amplification (and the associated frequency) in several deformation scenarios and to compare it against the maximum amplification achieved for Kerr black holes. 

 This work is organized as follows. In Sec.~II, we introduce the Kerr-like black-bounce metric, discuss the associated spacetime structure, and examine the Klein-Gordon equation that governs the propagation of scalar fields. In Sec.~III, we outline the numerical scheme employed to solve the wave equation and explore the dependence of the superradiant amplification factor on the parameters that characterize deviations from the Kerr geometry. Finally, Sec.~IV is dedicated to our concluding remarks. Throughout this work we use $G = c = \hbar = 1$ units. 


\section{Kerr-like black-bounce spacetimes} 
\label{Sec:def_Kerr}
We consider the rotating regular black hole geometry obtained in Ref.~\cite{Xu:2021lff} by applying the Newman-Janis algorithm to a static black-bounce metric. The spacetime, which generalizes the rotating Simpson-Visser solution~\cite{Simpson_2019,Mazza:2021rgq}, is characterized by the line element
\begin{align}
	ds^{2}&=-\left[1-\frac{2M(r^2+p^2)r^k}{(r^{2n}+p^{2n})^{\frac{k+1}{2n}} \Sigma}\right] \, dt^{2} +  \frac{\Sigma}{\Delta} \,  dr^{2}  \nonumber \\ & + \Sigma \, d\theta ^{2}  - \frac{4a \sin^{2} \theta}{\Sigma} \frac{M(r^2 + p^2)r^k}{(r^{2n} + p^{2n})^{\frac{k+1}{2n}}}  \, dtd\phi  \nonumber \\ &+ \frac{\sin ^{2}\theta}{\Sigma} \left[ (r^2 + p^2 + a^2)^2 - a^2 \Delta \sin ^{2}\theta \right] \, d\phi ^{2},	\label{eq:full_deformed_Kerr_line_element}
\end{align}
with
\begin{align}
	\Delta &= r^2 + p^2 + a^2 - \frac{2M(r^2+p^2)r^k}{(r^{2n}+p^{2n})^{\frac{k+1}{2n}}} ,  \\ 
	\Sigma &= r^2 + p^2 + a^2 \cos^2 \theta.
	 \end{align}

Here, $M$ is the mass of the spacetime, $a$ is the specific angular momentum, and $p$ is the black-bounce regularization parameter. The integers $k \geq 0$ and $n \geq 1$ are referred to as the deformation parameters. The choice $(k,n) = (0,1)$ yields the rotating Simpson-Visser spacetime, while $(k,n) = (2,1)$ corresponds to the rotating Bardeen-like spacetime~\cite{Xu:2021lff,Lobo:2020ffi}. Note that, regardless of the deformation exponents $k$ and $n$, the spacetime reduces to the Kerr spacetime when $p = 0$. In contrast, a nonvanishing parameter $p$ removes the central singularity and, together with $(k,n)$, controls the deviation from Kerr. In particular, the spacetime corresponds to a traversable wormhole when $p > p_c$ and to a regular black hole when $0 < p < p_c$, where the critical black-bounce parameter $p_c$ depends on the spin as well as on the deformation exponents~\cite{Xu:2021lff}. 

\begin{figure*}[!htbp]
	\centering
	\includegraphics[width = 1 \linewidth]{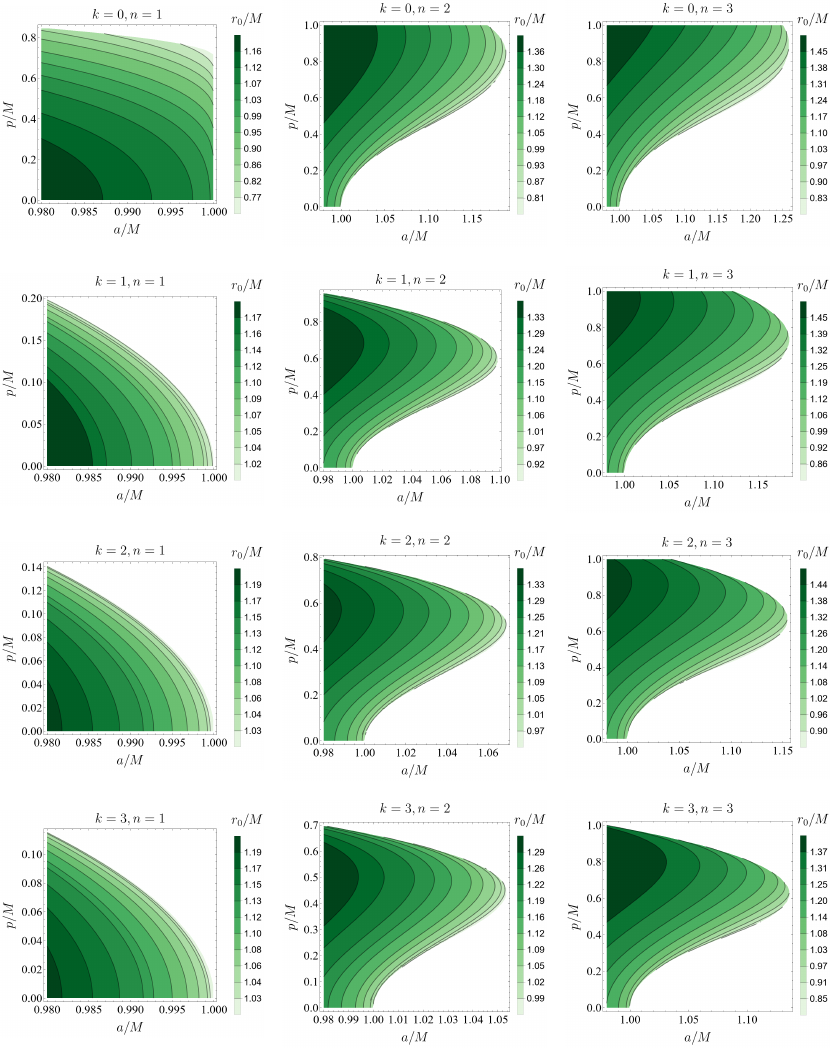}
	\caption{Contour plots of the event horizon radius $r_0/M$ as a function of the parameters $a/M$ and $p/M$ for different values of the deformation exponents $k$ and $n$. Each column corresponds to a given $n \in \{1, 2, 3\}$, while each row corresponds to a given $k \in \{ 0, 1, 2, 3\}$. The panels highlight how the interplay between spin, black-bounce parameter and deformation exponents constrains the parameter space of black hole configurations.}
	\label{fig:r0} 	
\end{figure*}

In Fig.~\ref{fig:r0}, we show the location $r_0$ of the event horizon for black holes described by the metric \eqref{eq:full_deformed_Kerr_line_element}. Each panel, corresponding to a particular choice of the pair $(k,n)$, exhibits the dependence of $r_{0}/M$ on both the dimensionless spin $a/M$ and the black-bounce parameter $p/M$. The white region in each panel corresponds to wormhole configurations (when $p>0$) and to naked singularities (when $p=0$).
 The green colored regions correspond to black hole configurations. 
 
We observe a clear distinction between the $n=1$ and the $n>1$ families. For $n=1$, the horizon radius $r_0/M$ decreases monotonically with increasing $a/M$ and $p/M$, for all values of $k$. Moreover, for fixed $(a/M,\,p/M)$, larger $k$ values lead to smaller black holes. For $n>1$, the dependence on spin is analogous: $r_0/M$ decreases with $a/M$. However, the dependence on the black-bounce parameter $p/M$ becomes non-monotonic. In particular, for each fixed $k$, the horizon radius initially increases with $p/M$, reaches a maximum at an intermediate value, and then decreases for larger $p/M$. 

\begin{figure*}[!htbp]
	\centering
	\includegraphics[width = 1 \linewidth]{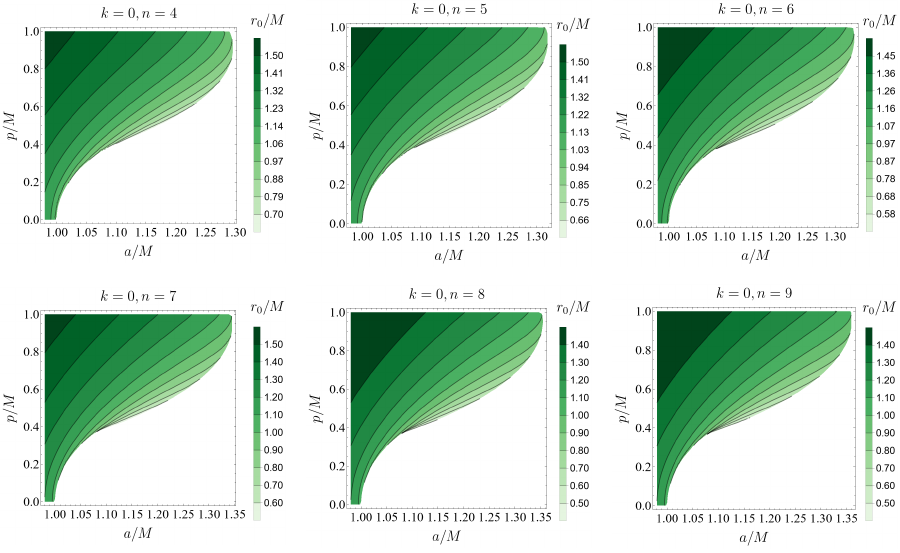}
	\caption{Contour plots of the event horizon radius $r_0/M$ as a function of the parameters $a/M$ and $p/M$ for the deformation exponents $k=0$ and $n \in \{4, \ldots, 9\}$.}
	\label{fig:r0_extra} 	
\end{figure*}

Regarding the role of the parameters $(k,n)$, we note that while $n=1$ does not admit ultra-spinning (i.e., $a/M > 1$) black hole configurations, the $n>1$ family does, as shown in the second and third columns of Figs.~\ref{fig:r0} and \ref{fig:r0_extra}. In particular, for the $n=2$ and $n=3$ configurations, the allowed range of black hole spins increases with $p/M$, up to a critical value at which the maximum admissible spin is reached. We also observe that, for fixed $n$, larger values of $k$ progressively restrict the black hole parameter space.

\subsection{Scalar field dynamics}

We are interested in the dynamics of a test massless scalar field $\Phi$ propagating in rotating regular black-bounce geometries described by \eqref{eq:full_deformed_Kerr_line_element}. The scalar field obeys the Klein-Gordon equation
\begin{equation} \label{eq:KG}
	\dfrac{1}{\sqrt{-g}}\partial _{\mu }\left(  g^{\mu \nu }\sqrt{-g}\partial _{\nu }\Phi \right) =0, 
\end{equation}
which can be separated due to the symmetries of the spacetime. In fact, given the stationarity and axisymmetry of the background, we decompose the scalar field into modes of frequency $\omega$, azimuthal number $m \in \mathbb{Z}$, and orbital number $\ell \ge m$ using the standard ansatz 
\begin{equation}
 \Phi_{\omega \ell m} (t, \phi, \theta, r) = e^{-i\omega t + im\phi} R_{\omega \ell m}(r) S_{\omega \ell m}(\theta).
 \label{eq:ansatz}
\end{equation}
This decomposition leads to decoupled radial and angular equations, involving a separation constant $\lambda_{\omega \ell m}$.
The angular part yields spheroidal harmonics satisfying
\begin{align}
	\frac{1}{\sin \theta} \frac{d}{d \theta} \left( \sin \theta \frac{dS_{\omega \ell m}}{d \theta} \right)  \nonumber &  +  \\ \bigg( \lambda_{\omega \ell m} -a^{2}\omega^2\cos ^{2}\theta   -  \dfrac{m^{2}}{\sin ^{2}\theta }\bigg) S_{\omega \ell m} & =0. \label{eq:angular_eq}
\end{align} 
The radial equation, on the other hand, takes the form
 \begin{align} \label{eq:radial_eq}
 	\Delta \frac{d}{dr}\left(\Delta \frac{d R_{\omega \ell m}}{dr} \right) & + \nonumber \\
 	\! \! \bigg[ \left(\omega \left( r^{2}+ p^{2} + a^{2}\right) - am \right)^{2}  -\lambda_{\omega \ell m} \Delta \bigg] R_{\omega \ell m} &=0.
 \end{align} 
 For simplicity, we shall omit the indices $\omega \ell m$ in $R_{\omega \ell m}$, $S_{\omega \ell m}$, and $\lambda_{\omega \ell m}$.

By defining the tortoise coordinate $r_*$ through 
\begin{equation} \label{eq:tortoise}
\frac{dr_*}{dr} = \frac{r^2+ p^2 + a^2}{\Delta}
 \end{equation}
 the radial equation can be transformed into a Schr\"odinger-like equation for the dependent function $\Psi$ given by
 \begin{equation} \label{eq:psi}
\Psi = \frac{R}{\sqrt{r^2+ p^2 + a^2}}.
 \end{equation}
In fact, by using \eqref{eq:tortoise} and \eqref{eq:psi} we can recast Eq.~\eqref{eq:radial_eq} as 
\begin{equation} \label{eq:schr}
\frac{d^2 \Psi}{dr_*^2} + (\omega^2 - V) \Psi = 0,
 \end{equation}  
 where the effective potential $V=V(\omega,r)$ is  
  \begin{align}
V=\frac{2am\omega}{r^2 + p^2 + a^2}
+\frac{\lambda\Delta-a^2m^2}{(r^2 + p^2 + a^2)^2} \nonumber \\
+\frac{\Delta\left(\Delta+r\Delta'\right)}{(r^2 + p^2 + a^2)^3}
-\frac{3r^2\Delta^2}{(r^2 + p^2 + a^2)^4}, \label{eq:effective} \end{align}
with primes denoting derivatives with respect to $r$.

The solution of the radial equation, which corresponds to ingoing waves in the near-horizon regime, is 
 \begin{equation} \label{eq:BC_horizon_psi}
	\Psi(r_*) \approx  e^{-i (\omega-m\Omega_H) r_*}, \quad r_* \rightarrow -\infty,
\end{equation}
or, equivalently,
 \begin{equation} \label{eq:BC_horizon}
	R\left( r\right) \approx \left( r-r_{0}\right) ^{ - \frac{i (a^2 + p^2 + r_{0}^2)}{\Delta'(r_{0})} \left(\omega  - m \Omega_H \right)}, \quad r \rightarrow r_0,
	\end{equation}
with \begin{equation}
 \Omega_H = \frac{a}{a^2 + p^2 + r_{0}^2}.
\end{equation}
In contrast, at spatial infinity, the asymptotic solution is
\begin{equation} \label{eq:aymptR_psi}
	 \Psi(r_*) \approx \mathcal{I} \, e^{-i\omega r_*} +  \mathcal{R} \, e^{i\omega r_*}, \quad r_* \rightarrow \infty,
\end{equation}
or, equivalently,
\begin{equation} \label{eq:aymptR}
	 R(r) \approx \mathcal{I} \, \frac{e^{-i\omega r}}{r^{1+2i M\omega}}  +  \mathcal{R} \, \frac{e^{i\omega r}}{r^{1-2i M\omega}}, \quad  r \rightarrow \infty,
\end{equation}
where $\mathcal{I}$ and $\mathcal{R}$ are related to the amplitudes of the incident and reflected waves, respectively. The superradiant amplification factor Z is defined as
\begin{equation} \label{eq:Zdef}
Z = Z(M\omega,p/M,a/M,k,n,\ell,m) = \frac{|\mathcal{R}|^{2}}{|\mathcal{I}|^{2}} - 1.
\end{equation}

\begin{figure*}[!htbp]
	\centering
	\includegraphics[width = 1 \linewidth]{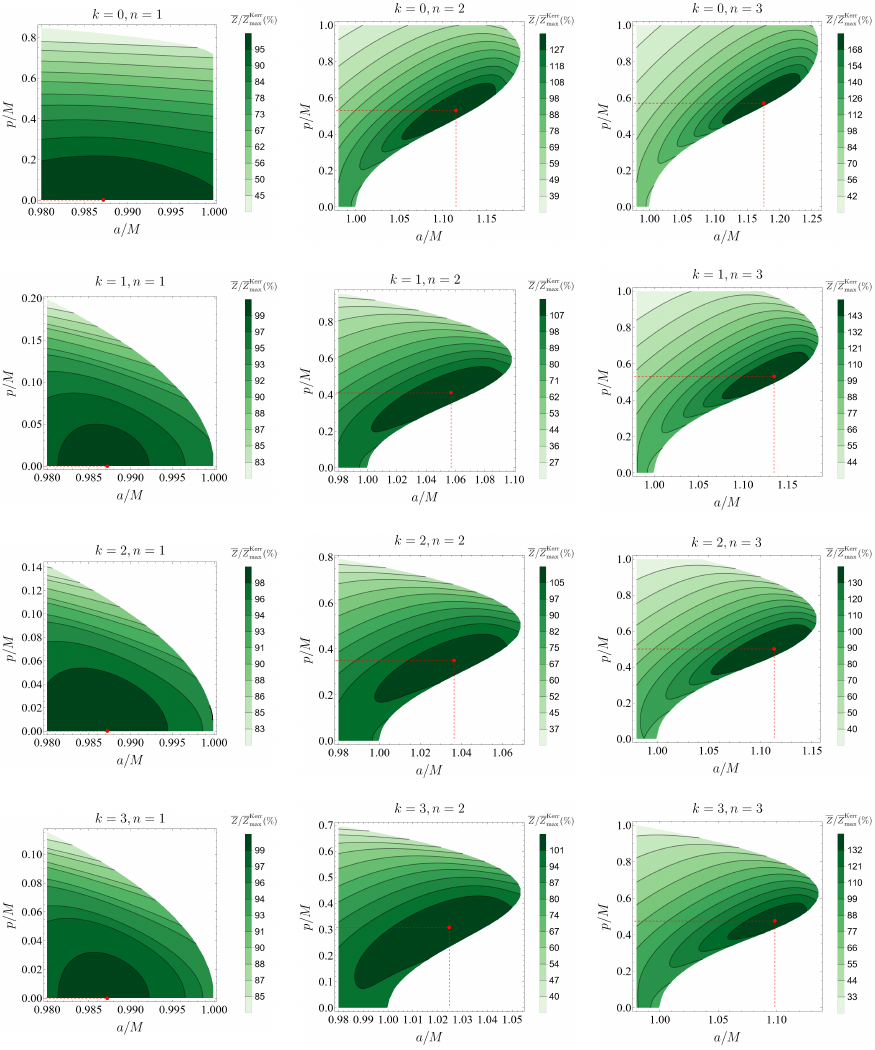}
	\caption{Contour plots of the relative maximum superradiant amplification $\overline{Z}/\overline{Z}^{\mathrm{Kerr}}_{\mathrm{max}}$ in the $(a/M,p/M)$ parameter space for different values of the deformation exponents $k$ and $n$. Each column corresponds to a fixed $n=1,2,3$, and each row to a fixed $k=0,1,2,3$. The color scale quantifies the enhancement or suppression of the amplification relative to the Kerr case, highlighting how deviations controlled by $p/M$ and the deformation exponents modify the efficiency of superradiant scattering. The red dot in each panel denotes the location of the global maximum $\overline{Z}_{\mathrm{max}}$.}
	\label{fig:Z} 	
\end{figure*}

\begin{figure*}[!htbp]
	\centering
	\includegraphics[width = 1 \linewidth]{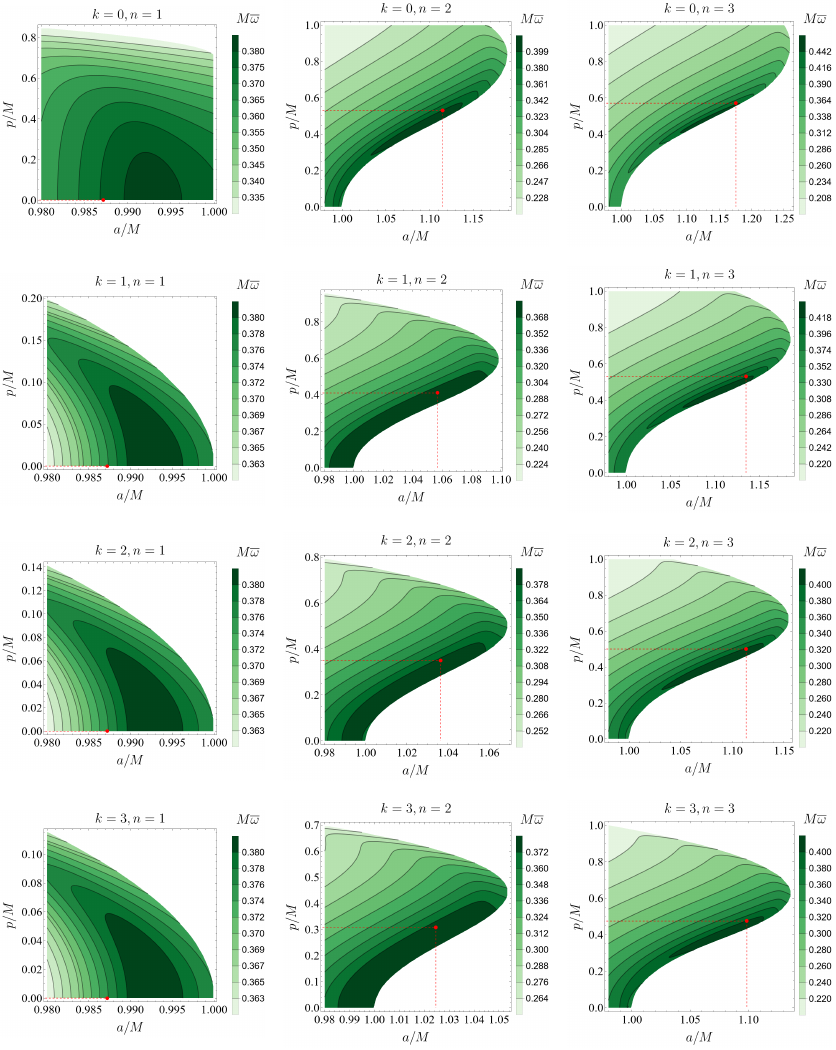}
	\caption{Contour plots of the frequency $M\overline{\omega}$ associated with the maximum superradiant amplification in the $(a/M,p/M)$ parameter space for different values of the deformation exponents $k$ and $n$. Each column corresponds to a fixed $n=1,2,3$, and each row to a fixed $k=0,1,2,3$. The color scale indicates the value of $M\overline{\omega}$, illustrating how the peak frequency shifts with the spin $a/M$, the black-bounce parameter $p/M$, and the deformation exponents. The red dot in each panel marks the location of the global maximum $\overline{Z}_{\mathrm{max}}$ shown in Fig.~\ref{fig:Z}.}
	\label{fig:omega} 	
\end{figure*}


\section{Amplification factors}

\subsection{Numerical scheme}
To compute the amplification factor $Z$, we solve the scattering problem using the following numerical procedure. For fixed deformation exponents $(k,n)$, and for each choice of spin $a/M$ and black-bounce parameter $p/M$, we first determine the roots of the metric function $\Delta(r)$ and identify the position $r_0$ of the event horizon. We then fix the mode parameters $(\ell,m)$ and the frequency $\omega$, and solve the radial wave equation \eqref{eq:radial_eq} by imposing initial conditions near the event horizon, specifically at $r_1 = r_0 + 10^{-8}$. The radial equation is integrated numerically up to $r_2 = 200 M$, where we extract the radial function $R$ and its derivative. Using $R(r_2)$ and $R'(r_2)$, we compute the incident and reflected parameters, respectively $\mathcal{I}$ and $\mathcal{R}$, via \eqref{eq:aymptR}, and hence obtain the amplification factor $Z$
from Eq.~\eqref{eq:Zdef}. Keeping all other parameters fixed, we then vary $\omega$ and repeat the numerical integration to determine the maximum amplification with respect to the frequency,
\begin{equation} \label{overZ}
\overline{Z} = \max \{ Z : M\omega \in \mathbb{R} \},
\end{equation}
together with the corresponding frequency $\overline{\omega}$.

To investigate the impact of deviations from the Kerr geometry on superradiance, we compute  $\overline{Z}$ and $\overline{\omega}$ as we scan the parameter space $(a/M,p/M)$. We focus on $(\ell,m)=(1,1)$ modes in the high spin regime, where superradiant amplification is strongest for Kerr black holes. The scan starts at \(a/M = 0.98\) and \(p/M = 0\), corresponding to a rapidly rotating Kerr black hole, for which the peak amplification is expected to occur at $M \omega \sim 0.38$~\cite{Brito2020}.
From this initial point, we increase the spin $a/M$ in steps $\Delta a/M$ and, at each step, verify whether an event horizon still exists. Whenever it does, we compute a new set of amplification factors $Z$ using a refined frequency interval centered on the peak amplification of the previous iteration:
\begin{equation}
M\omega \in [M\overline{\omega}_i - 0.03, \, M\overline{\omega}_i + 0.03],
\end{equation}
where $M\overline{\omega}_i$ denotes the peak frequency obtained for the preceding spin value. From this dataset, we extract the peak amplification $\overline{Z}$ and the corresponding frequency $\overline{\omega}$ at the current spin. For each fixed $a/M$, we then perform a scan in the $p$-direction, increasing the black-bounce parameter from $p/M = 0$ up to $p/M = 1$ in steps of $\Delta p/M$. Using an analogous refinement procedure, we determine $\overline{Z}$ and $\overline{\omega}$ for each pair $(a/M, p/M)$ in the parameter space. The increments $\Delta a/M$ and $\Delta p/M$ are chosen to ensure adequate resolution in the resulting figures.

Once the full $(a/M, p/M)$ domain of interest has been scanned, we identify the maximum superradiant amplification factor $\overline{Z}_{\mathrm{max}}$ across the parameter space for the chosen pairs $(\ell,m)$ and $(k,n)$,
\begin{equation}
\overline{Z}_{\mathrm{max}} = \max \{\overline{Z} : a/M \ge 0.98, \, p/M \ge 0 \}.
\end{equation}
We denote the corresponding frequency by $\overline{\omega}_{\mathrm{max}}$, and the location in parameter space where this maximum is attained by $(a_{\mathrm{max}}/M, p_{\mathrm{max}}/M)$. To improve the precision of our investigation, we perform a secondary, high-resolution scan in the region surrounding this initial point. Because the amplification profile varies with each $(k,n)$ configuration, we employ a case-specific refinement grid rather than fixed increments. The  step sizes (i.e.,\ $\Delta a/M$ and $\Delta p/M$) and the scan boundaries are individually adapted for each configuration to fully resolve the local maximum. This targeted high-resolution scan is then used to refine the values of $\overline{Z}_{\mathrm{max}}$, $M\overline{\omega}_{\mathrm{max}}$, $a_{\mathrm{max}}/M$, and $p_{\mathrm{max}}/M$.


\subsection{Numerical results}

Our main results are presented in Figs.~\ref{fig:Z} and \ref{fig:omega}.  We consider twelve representative cases: all combinations $(k,n)$ satisfying $0 \leq k \leq 3$ and $1 \leq n \leq 3$. Fig.~\ref{fig:Z} displays the maximum amplification factor $\overline{Z}$ at each point $(a/M, p/M)$, normalized to the Kerr global maximum $\overline{Z}^{\mathrm{Kerr}}_{\mathrm{max}} \simeq 0.003674$, which occurs at $a/M \approx 0.9872$~\cite{Brito2020}. Fig.~\ref{fig:omega} shows the frequency $M\overline{\omega}$ associated with the peak amplification at each point in parameter space. In both figures, the red dot marks the location $(a_{\mathrm{max}}/M, p_{\mathrm{max}}/M)$ where the maximum amplification within each panel, i.e.~$\overline{Z}_{\mathrm{max}}$, is attained. As seen in Fig.~\ref{fig:omega}, the frequency $\overline{\omega}_{\mathrm{max}}$ corresponding to $\overline{Z}_{\mathrm{max}}$ in each panel does not coincide with the global maximum of $\overline{\omega}$.

The first column of Figs.~\ref{fig:Z} and \ref{fig:omega} corresponds to the case $n=1$, namely the configurations $(k,n) = (0,1)$, $(1,1)$, $(2,1)$, and $(3,1)$. From Fig.~\ref{fig:Z} we observe that, for $n=1$, the peak amplification $\overline{Z}$ decreases monotonically with increasing $p/M$ at fixed $a/M$. Conversely, for sufficiently small values of $p/M$, the dependence of the peak amplification on the spin is nonmonotonic: $\overline{Z}$ initially increases with $a/M$, reaches a local maximum, and then decreases as $a/M \to 1$. In particular, we verify that superradiance is maximized at $p/M = 0$ and $a/M = 0.9872$, which coincides with the spin value that maximizes superradiance in the Kerr spacetime. As a consequence, for $n=1$, the superradiant amplification of regular rotating black-bounce spacetimes never exceeds the Kerr maximum. In other words, the largest amplification is always achieved in the Kerr limit, and the black-bounce deformation does not enhance superradiance. 
The corresponding peak frequencies are shown in the first column of Fig.~\ref{fig:omega}. For sufficiently small $p/M$, the peak frequency exhibits a nonmonotonic dependence on the spin, increasing with $a/M$ before decreasing as the extremal limit is approached.

\begin{table}[htbp!]
\centering
\begin{tabular}{cccccc}
\hline\hline
$(k,n)$ & $a_{\mathrm{max}}/M$ & $p_{\mathrm{max}}/M$ & $M \overline{\omega}_{\mathrm{max}}$ & $\overline{Z}_{\mathrm{max}}$ & $\overline{Z}_{\mathrm{max}}/\overline{Z}_{\mathrm{max}}^{\mathrm{Kerr}}$ \\ 
\hline 
(0,2) & 1.115 & 0.53 & 0.4003 & 0.005236 & 142\% \\ 
(1,2) & 1.057 & 0.41 & 0.3868 & 0.004425 & 120\% \\ 
(2,2) & 1.036 & 0.35 & 0.3830 & 0.004173 & 114\% \\ 
(3,2) & 1.025 & 0.31 & 0.3810 & 0.004048 & 110\% \\ 
(0,3) & 1.176 & 0.57 & 0.4351 & 0.007273 & 198\% \\ 
(1,3) & 1.135 & 0.53 & 0.4130 & 0.006037 & 164\% \\ 
(2,3) & 1.113 & 0.50  & 0.4049 & 0.005553 & 151\% \\ 
(3,3) & 1.099 & 0.47 & 0.4011 & 0.005279 & 144\% \\ 
\hline\hline 
\end{tabular}
\caption{Global maxima of the superradiant amplification factor for deformation exponents $n=2$ and $n=3$, with $0 \leq k \leq 3$. For each configuration, we list the maximum amplification factor $\overline{Z}_{\mathrm{max}}$ and its ratio relative to the Kerr value $\overline{Z}^{\mathrm{Kerr}}_{\mathrm{max}}$, together with the corresponding values of the spin parameter $a_{\mathrm{max}}/M$, the black-bounce parameter $p_{\mathrm{max}}/M$, and the frequency $M \overline{\omega}_{\mathrm{max}}$ at which the maximum occurs.}
\label{tab:table}
\end{table}

\begin{figure*}[!htbp]
	\centering
	\includegraphics[width = 0.96 \linewidth]{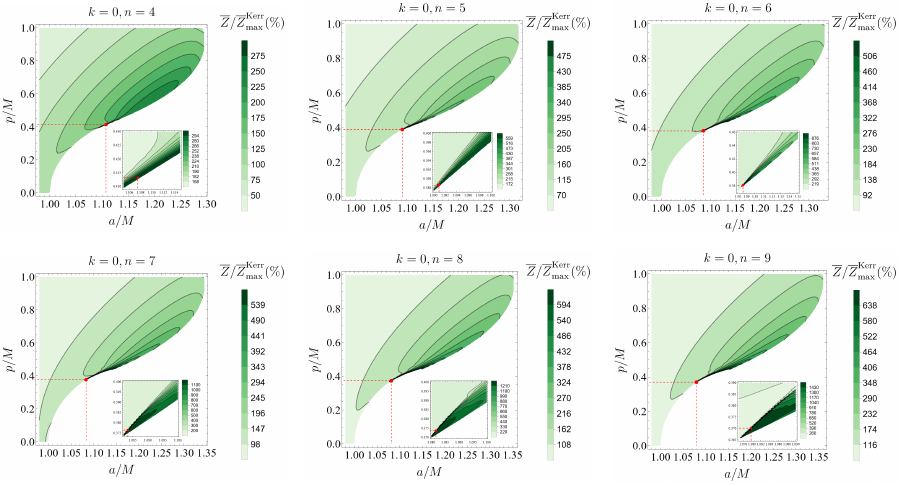}
	\caption{Contour plots of the relative maximum superradiant amplification $\overline{Z}/\overline{Z}^{\mathrm{Kerr}}_{\mathrm{max}}$ in the $(a/M,p/M)$ parameter space for the deformation parameters $k=0$ and $n = 4,\ldots,9$. The red dot in each panel denotes the location of the global maximum $\overline{Z}_{\mathrm{max}}$. The insets highlight the region of parameter space where a second peak emerges in the amplification spectrum and grows rapidly as the parameters are varied (see main text).}
	\label{fig:Z_extra} 	
\end{figure*}
\begin{figure*}[!htbp]
	\centering
	\includegraphics[width = 0.92 \linewidth]{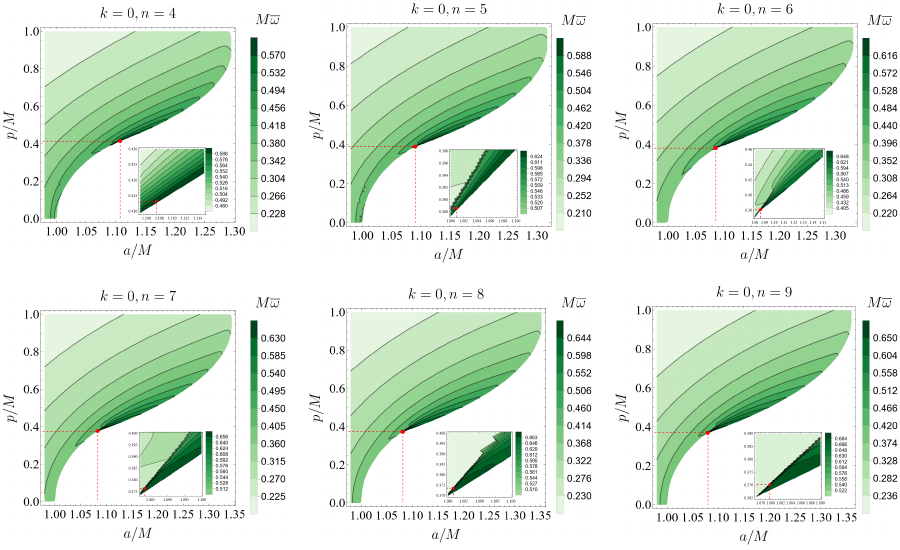}
	\caption{Contour plots of the frequency $M\overline{\omega}$ associated with the maximum superradiant amplification in the $(a/M,p/M)$ parameter space for the deformation parameters $k=0$ and $n \in \{4,\ldots,9\}$. The red dot in each panel marks the location of the global maximum $\overline{Z}_{\mathrm{max}}$ shown in Fig.~\ref{fig:Z_extra}. The insets highlight the region of parameter space where a second peak emerges in the amplification spectrum and grows rapidly as the parameters are varied (see main text).}
	\label{fig:omega_extra} 	
\end{figure*}

Next, we examine $n > 1$ configurations, specifically $n=2$ and $n=3$, which allow for superspinning $a/M>1$ black holes.  The results are presented in the second and third columns of Figs.~\ref{fig:Z} and \ref{fig:omega}.  In contrast to the $n=1$ case, we find regions of the parameter space $(a/M, p/M)$ where the superradiant amplification exceeds the maximum amplification for Kerr black holes. Table~\ref{tab:table} shows the parameter values corresponding to the global maxima of superradiance in each of the $n > 1$ configurations. For instance, when $(k,n)=(0,3)$, we find that the amplification factor can be as high as 198$\%$ of the Kerr result, indicating that black-bounce deformations can significantly enhance superradiant effects. 

To isolate the effects of the deformation exponents $k$ and $n$, we consider variations at fixed $n$ and at fixed $k$. Examining a given row of Fig.~\ref{fig:Z} allows us to assess the effect of increasing $n$ for fixed values of $k = 0, 1, 2,$ and $3$. We find that increasing $n$ not only shifts the location of the amplification peak in parameter space towards larger $a/M$ and $p/M$, but also leads to an enhancement of the superradiant amplification relative to the Kerr case. Conversely, examining a fixed column of Fig.~\ref{fig:Z} highlights the effect of increasing $k$ for fixed $n = 1, 2,$ and $3$. For $n=1$, increasing $k$ from $0$ to $3$ produces no change in the maximum amplification, since the superradiant peak always coincides with the Kerr limit. For $n=2$ and $n=3$, increasing $k$ tends to suppress the strength of superradiance. Table~\ref{tab:table} shows that increasing $k$ from 0 to 3 reduces the enhancement of superradiance from $142\%$ to $110\%$ when $n=2$. Similarly, for $n=3$, the enhancement is reduced from $198\%$ at $k=0$ to $144\%$ at $k=3$. 

\begin{figure}[!htbp]
	\centering
	\includegraphics[width = \columnwidth]{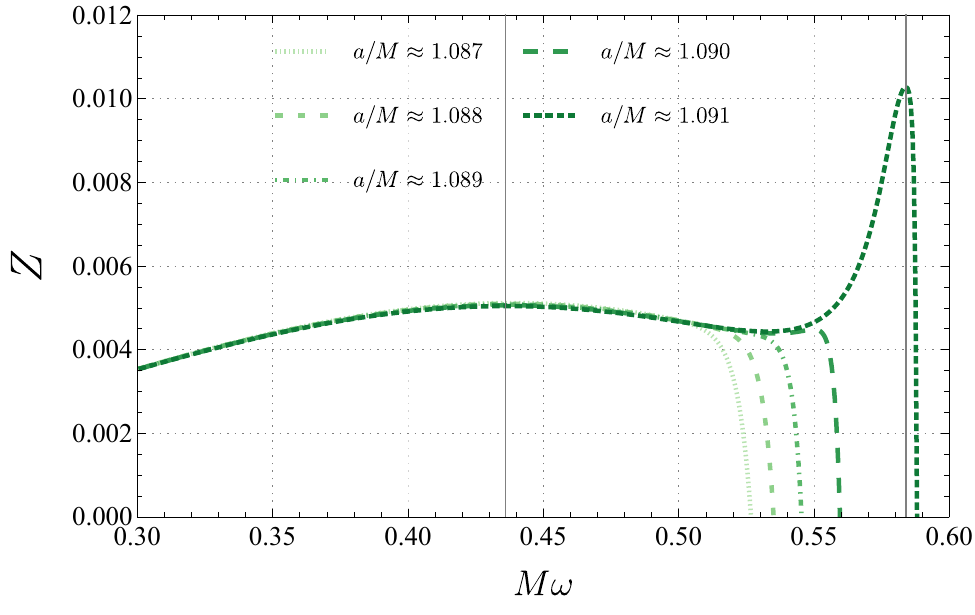}
	\caption{Amplification spectra for $(k,n)=(0,5)$, $p/M=0.389$, and $\ell=m=1$ for different values of $a/M$. As the spin increases, a second local maximum of the amplification factor $Z$ develops and grows sharply with $a/M$. The vertical grey lines indicate the locations of the two local maxima for $a/M=1.091$.}
	\label{fig:double_peak} 	
\end{figure}

To further investigate the superradiant enhancement effect as $n$ increases, we fix $k=0$ and compute the maximum amplification factor $\overline{Z}$, together with the corresponding frequency $M\overline{\omega}$, for $n = 4,\ldots,9$. The results are presented in Figs.~\ref{fig:Z_extra} and \ref{fig:omega_extra}. We observe that the location of the global maximum $\overline{Z}_{\mathrm{max}}$ appears to converge towards the boundary of the green region separating black hole spacetimes from wormhole configurations. The values of $\overline{Z}_{\mathrm{max}}$ and the corresponding frequencies $M \overline{\omega}_{\mathrm{max}}$ are listed in Table~\ref{tab:table_extra}. The amplification factors exhibit a steep increase with $n$. An important factor contributing to this enhancement is the emergence of a second peak in the amplification spectrum for the configurations with $k=0$ and $n \geq 4$. We illustrate the emergence of this peak, for the particular case $(k,n)=(0,5)$, in Fig.~\ref{fig:double_peak}. Fixing $p/M=0.389$ and $\ell=m=1$, we show the amplification spectra for different values of $a/M$, with $a/M$ increasing towards the boundary of the parameter space. For $a/M \gtrsim 1.090$, a second local maximum emerges at a higher frequency and grows sharply as $a/M$ increases. For example, at $a/M=1.091$, the amplification spectrum exhibits local maxima at $M\omega=0.436$ and $M\omega=0.584$, whose locations are indicated by the solid vertical grey lines in Fig.~\ref{fig:double_peak}. The rapid growth of this second peak is responsible for the darkest patches in Figs.~\ref{fig:Z_extra} and \ref{fig:omega_extra}, highlighted in insets of the respective panels, where the contour lines become concentrated and the maximum amplification increases sharply. 

\begin{table}[htbp!]
\centering
\begin{tabular}{cccccc}
\hline\hline
$(k,n)$ & $a_{\mathrm{max}}/M$ & $p_{\mathrm{max}}/M$ & $M \overline{\omega}_{\mathrm{max}}$ & $\overline{Z}_{\mathrm{max}}$ & $\overline{Z}_{\mathrm{max}}/\overline{Z}_{\mathrm{max}}^{\mathrm{Kerr}}$ \\ 
\hline 
(0,4) & 1.1075 & 0.4132 & 0.6034 & 0.01158 & 315\% \\ 
(0,5) & 1.0909 & 0.3886 & 0.6419 & 0.02263 & 616\% \\ 
(0,6) & 1.0860 & 0.3807 & 0.6698 & 0.03457 & 941\% \\ 
(0,7) & 1.0834 & 0.3764 & 0.6785 & 0.04546 & 1237\% \\ 
(0,8) & 1.0817 & 0.3735 & 0.6846 & 0.05320 & 1448\% \\ 
(0,9) & 1.0798 & 0.3701 & 0.6918 & 0.05924 & 1612\% \\
\hline\hline 
\end{tabular}
\caption{Global maxima of the superradiant amplification factor for deformation exponents $k=0$ and $n=4,\dots 9$. For each configuration, we list the maximum amplification factor $\overline{Z}_{\mathrm{max}}$ and its ratio relative to the Kerr value $\overline{Z}^{\mathrm{Kerr}}_{\mathrm{max}}$, together with the corresponding values of the spin parameter $a_{\mathrm{max}}/M$, the black-bounce parameter $p_{\mathrm{max}}/M$, and the frequency $M \overline{\omega}_{\mathrm{max}}$ at which the maximum occurs.}
\label{tab:table_extra}
\end{table}

\begin{figure}[!htbp]
	\centering
	\includegraphics[width = \linewidth]{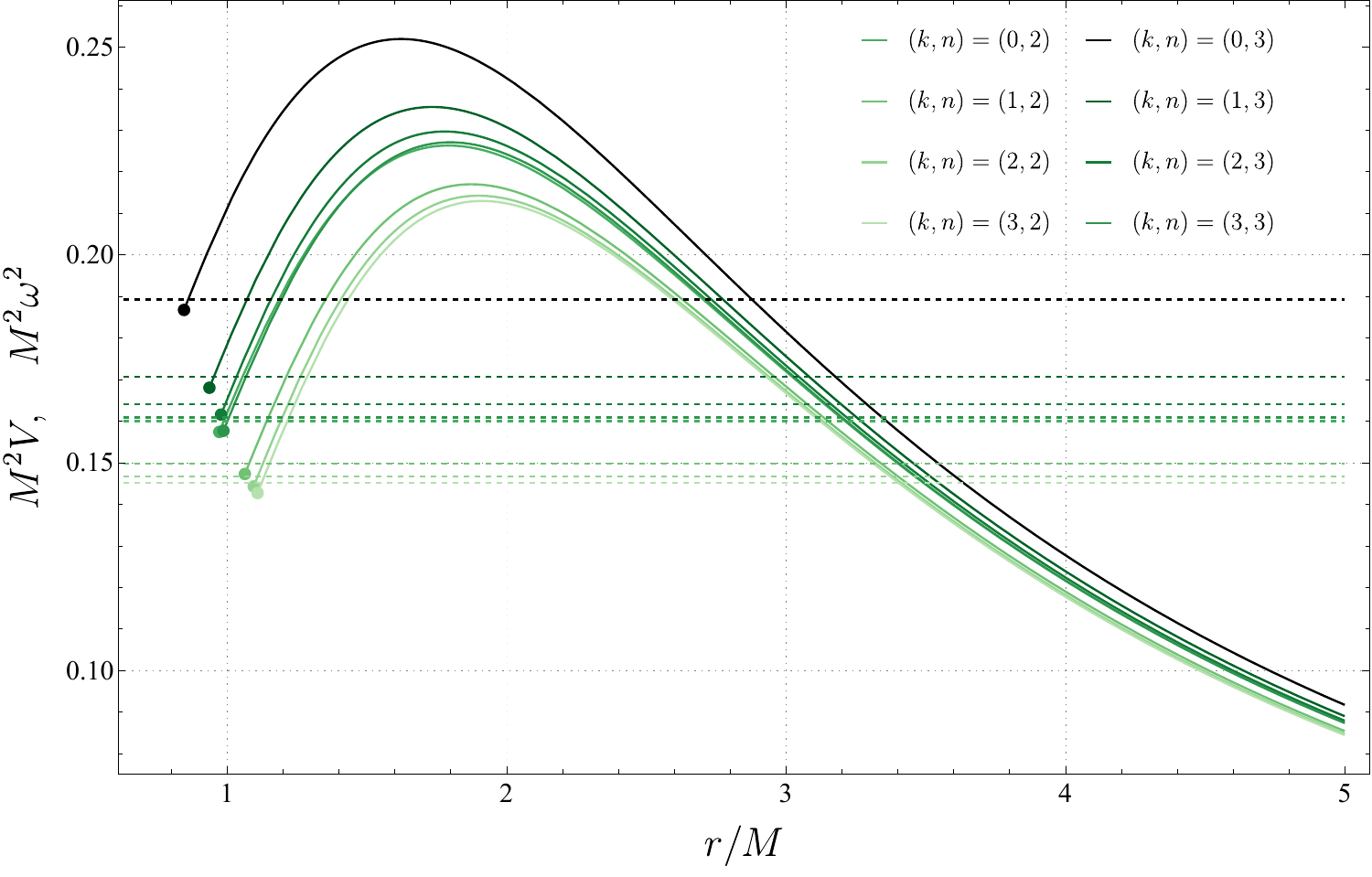}
	\caption{Effective potential $V=V(\omega,r)$ for the parameters corresponding to the global maxima shown in Table \ref{tab:table}. The horizontal (dashed) lines correspond to $\omega_{\mathrm{max}}^2$, which compares to the effective potential according to Eq.~\eqref{eq:schr}. The dots mark the value of the potential at the corresponding event horizon. Note that the effective potential and the effective energy $\omega^2$ increase together and, as they increase, the global maximum $\overline{Z}_{\mathrm{max}}$ also increases. In particular, the highest barrier and the largest $\overline{Z}_{\mathrm{max}}$ correspond to $(k,n)=(0,3)$, while the lowest barrier and lowest $\overline{Z}_{\mathrm{max}}$ correspond to $(k,n)=(3,2)$.}
	\label{fig:pot_tab1} 	
\end{figure}

\begin{figure}[!htbp]
	\centering
	\includegraphics[width = \linewidth]{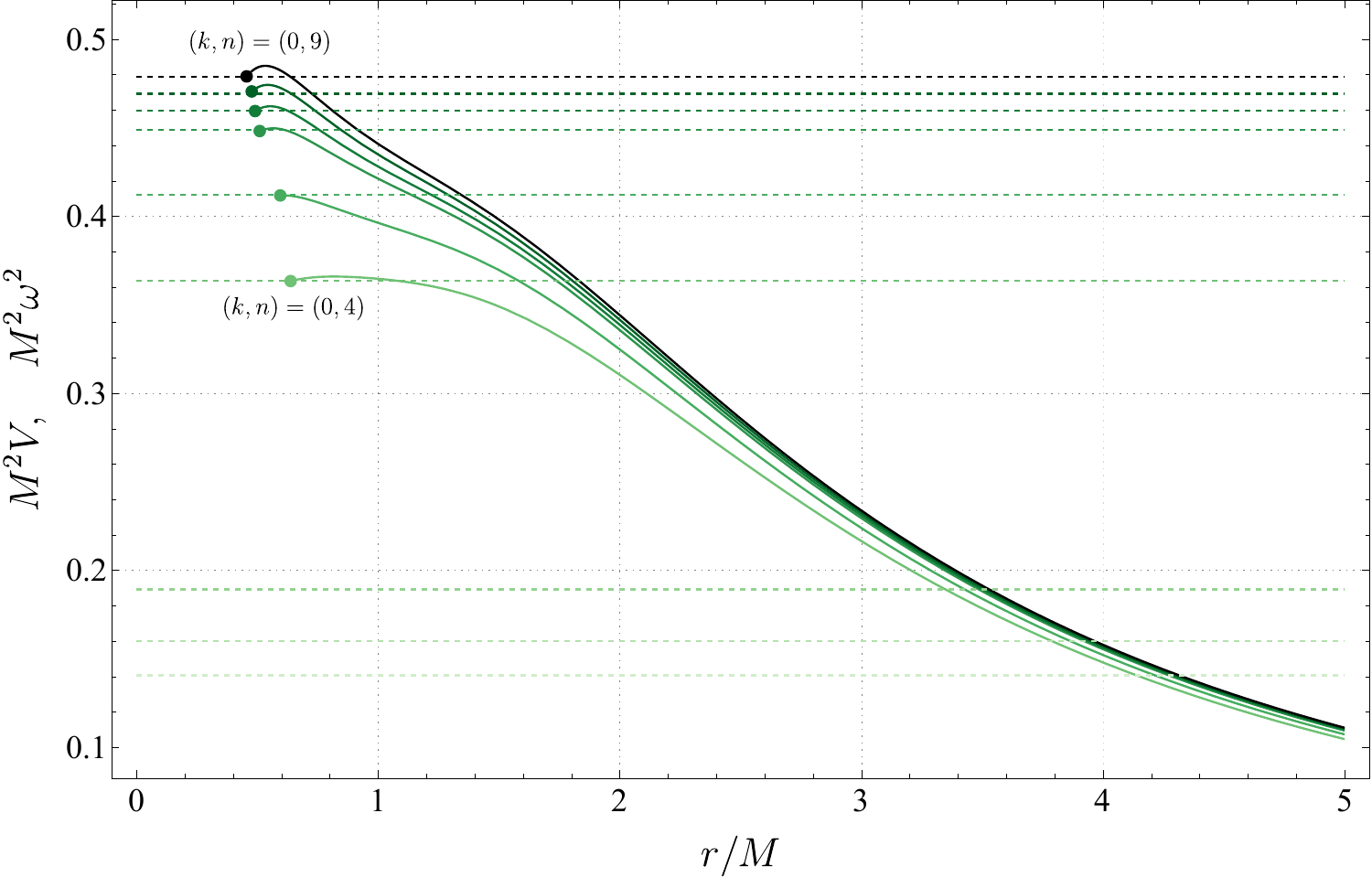}
	\caption{Effective potential $V=V(\omega,r)$ for the parameters corresponding to the global maxima shown in Table \ref{tab:table_extra}. The horizontal (dashed) lines correspond to $\omega_{\mathrm{max}}^2$, which compares to the effective potential according to Eq.~\eqref{eq:schr}. The dots mark the value of the potential at the corresponding event horizon. 
    As $n$ increases, both the effective potential and the effective energy $\omega^2$ increase and, as they increase, the global maximum $\overline{Z}_{\mathrm{max}}$ also increases. In particular, the highest barrier and the largest $\overline{Z}_{\mathrm{max}}$ correspond to $(k,n)=(0,9)$, while the lowest barrier and lowest $\overline{Z}_{\mathrm{max}}$ correspond to $(k,n)=(0,4)$.}
	\label{fig:pot_tab2} 	
\end{figure}

The role of the deformation exponents $k$ and $n$ on superradiance can be qualitatively understood in terms of the effective potential \eqref{eq:effective}. In Fig.~\ref{fig:pot_tab1}  we plot the effective potential $V$ and the frequency-squared $\omega^2$, which we refer to as the effective wave energy in view of Eq.~\eqref{eq:schr}, for the parameters listed in Table~\ref{tab:table}. For fixed $n$, the configurations corresponding to the maximum amplification exhibit progressively lower potential barriers and effective wave energies as $k$ increases. This simultaneous decrease is consistent with a reduction in the amplitude $\mathcal{R}$ of the reflected wave relative to the amplitude $\mathcal{I}$ of the incident wave, and hence with a decrease in $\overline{Z}_{\mathrm{max}}$ and a suppression of superradiant amplification. In contrast, for fixed $k$, increasing the deformation exponent $n$ raises both the potential barrier and the corresponding effective wave energy. This behaviour is consistent with a larger fraction of the incident wave being reflected by the potential barrier, thereby enhancing the amplification factor $\overline{Z}_{\mathrm{max}}$. The effect of increasing $n$ at fixed $k$ is further illustrated in Fig.~\ref{fig:pot_tab2}, which shows the effective potential and effective wave energy for the parameters listed in Table~\ref{tab:table_extra}.

%



\section{Final remarks}

In this work, we investigated the superradiant scattering of massless scalar fields by a rotating black-bounce spacetime, a regular geometry that extends the Simpson–Visser solution through a set of deformation parameters. By varying the pair $(k,n)$ characterizing the metric, together with the black-bounce parameter $p/M$ and the dimensionless spin $a/M$, we mapped the behavior of the superradiant amplification factor over a broad region of the parameter space. Our results show that certain combinations of parameters lead to substantial deviations from the Kerr black hole scenario. 
In particular, we identified regimes, specially for higher values of $n$, in which the superradiant amplification significantly exceeds the maximum found for standard Kerr black holes.
These enhancements appear to be associated with modifications of the near-horizon geometry induced by the regularity structure which allow for superspinning configurations with $a/M>1$. 

An important outcome of our analysis is the identification of a nontrivial dependence of the peak amplification on the black-bounce parameter $p/M$. In several cases, the amplification exhibits a nonmonotonic behavior as $p/M$ increases, indicating the existence of optimal configurations for energy extraction. We were also able to investigate the roles of the deformation exponents $k$ and $n$. We find that the parameter $n$ plays a central role in enabling ultra-spinning configurations as $p/M$ increases, with larger $n$ leading to enhanced superradiant amplification. In contrast, increasing $k$ tends to suppress superradiance. This dependence on the deformation exponents can be qualitatively understood in terms of the interplay between the effective potential and the effective wave energy. For the configurations corresponding to maximum amplification, increasing $k$ is accompanied by a decrease in both the potential barrier $V$ and the effective wave energy $\omega^2$, a behaviour consistent with the suppression of superradiance. Conversely, increasing $n$ leads to higher values of both $V$ and $\omega^2$, a behaviour consistent with enhanced superradiant amplification.

Although our focus has been restricted to massless scalar fields, which provide a useful toy model for probing aspects of gravitational-wave physics, the  analysis developed here opens avenues for future investigation. 
Natural extensions include the study of electromagnetic and gravitational perturbations, as well as stability analyses in the ultra-spinning regime. Additional insights could be obtained through an analytical treatment in the low-frequency regime $(M\omega \ll 1)$, along the lines of the calculations performed, e.g., in~\cite{Starobinskil:1974nkd,Page:1976df,Richartz:2011vf}.
One can also ask whether the enhancement of superradiance persists in the presence of massive scalar fields, a situation in which superradiant scattering is typically accompanied by quasibound states and instabilities~\cite{Siqueira:2022tbc,Dolan:2012yt,Furuhashi:2004jk,Cardoso:2005vk,Hod:2012px}. Such instabilities may arise spontaneously in astrophysical environments, particularly when the compact object is surrounded by a massive bosonic field~\cite{Press:1972zz,Hod:2012px}, potentially leading to the formation of long-lived bosonic clouds that emit nearly monochromatic gravitational waves. The detection of such signals would provide indirect evidence for ultralight bosons and place constraints on their masses~\cite{Brito:2014wla}. A detailed analysis of the spectrum and lifetimes of these states could therefore shed light on the formation of bosonic clouds around regular rotating geometries, constrain the parameter space of black-bounce spacetimes, and assess their viability as beyond-Kerr alternatives in the era of multi-messenger astronomy. Finally, the role of exceptional points and their associated phenomenology, an area that has attracted significant attention recently~\cite{Motohashi:2024fwt,Cavalcante:2024swt,Cavalcante:2024kmy,Yang:2025dbn,Lo:2025njp,Cavalcante:2025abr}, constitutes another promising direction for future work.

\section*{Data Availability}

Data supporting the findings of this work are openly available~\cite{databb}.


\acknowledgments

This research was partially financed by the Coordena\c{c}\~ao de Aperfei\c{c}oamento de Pessoal de N\'{i}vel Superior (CAPES, Brazil) - Finance Code 001. PHCS acknowledges support from the São Paulo Research Foundation (FAPESP, Brazil), grant 2022/07298-4. MR acknowledges partial support from the Conselho Nacional de Desenvolvimento Científico e Tecnológico (CNPq, Brazil, grant 315991/2023-2), and from the São Paulo Research Foundation (FAPESP, Brazil, grants 2022/08335-0 and 2024/00923-6). PHCS thanks the Strong Group at the Niels Bohr Institute for their kind hospitality while part of this work was carried out. The Tycho supercomputer hosted at the SCIENCE HPC center at the University of Copenhagen was used for supporting this work.


\bibliography{ref}

\end{document}